\documentclass[aps,pra,showpacs,twocolumn,floatfix,reprint]{revtex4-1}
\usepackage{eurosym}
\usepackage{amsfonts}
\usepackage{amssymb}
\usepackage{amsmath}
\usepackage{amsbsy}
\usepackage{latexsym,epsfig,graphicx}
\usepackage{dcolumn}
\usepackage{bm}
\usepackage{graphicx}
\usepackage{subfigure}
\usepackage{color}
\usepackage{lipsum}
\usepackage[colorlinks,urlcolor=blue,citecolor=blue]{hyperref}

\begin{document}

\title{Dynamical spin-density waves in a spin-orbit-coupled Bose-Einstein
condensate}
\author{Yan Li$^{1,2}$}
\author{Chunlei Qu$^{2}$}
\author{Yongsheng Zhang$^{3,4,2}$}
\author{Chuanwei Zhang$^{2}$}
\thanks{Corresponding author.\\
Email: chuanwei.zhang@utdallas.edu}

\begin{abstract}
Synthetic spin-orbit (SO) coupling, an important ingredient for quantum
simulation of many exotic condensed matter physics, has recently attracted
considerable attention. The static and dynamic properties of a SO coupled
Bose-Einstein condensate (BEC) have been extensively studied in both theory
and experiment. Here we numerically investigate the generation and
propagation of a \textit{dynamical} spin-density wave (SDW) in a SO coupled
BEC using a fast moving Gaussian-shaped barrier. We find that the SDW
wavelength is sensitive to the barrier's velocity while varies slightly with
the barrier's peak potential or width. We qualitatively explain the
generation of SDW by considering a rectangular barrier in a one dimensional
system. Our results may motivate future experimental and theoretical
investigations of rich dynamics in the SO coupled BEC induced by a moving
barrier.
\end{abstract}
\affiliation{$^{1}$Department of Physics, East China Normal University, Shanghai 200241,
China\\
$^{2}$Department of Physics, University of Texas at Dallas, Richardson,
Texas 75080, USA\\
$^{3}$Laboratory of Quantum Information, University of Science and
Technology of China, Hefei 230026, China\\
$^{4}$Synergetic Innovation Center of Quantum Information and Quantum
Physics, University of Science and Technology of China, Hefei 230026, China }
\maketitle

\section{Introduction}

\label{Sec:Intro}

Spin-orbit (SO) coupling plays an important role for the emergence of many
exotic quantum phenomena in condensed matter physics \cite{Kane,Qi}. In this
context, the recent experimental realization of SO coupled neutral atoms
provides an excellent platform for the quantum simulation of condensed
matter phenomena because of the high controllability and free of disorder~%
\cite{Lin2011,Wang2012,Cheuk2012,Williams2013} of cold atoms. By dressing
two atomic internal states through a pair of lasers, a Bose-Einstein
condensate (BEC) with equal Rashba and Dresselhaus SO coupling has been
achieved~\cite{Lin2011,Fu2011,Pan2012,Qu2013,Hamner2014,Olson2014,Karina2014}%
. The static and dynamic properties of such SO coupled BEC~\cite%
{Dalibard2011,Wang2010,Wu2011,Ho2011,Zhang2012,Hu2012,Ozawa2012,Li2012,Xu2013,Goldman2013,Flatband}
have also been investigated. Notable experimental progress in SO coupled
BECs includes the observation of spin Hall effects~\cite{SpinHall} and
Dicke-type phase transition~\cite{Hamner2014}, the study of collective
excitations such as the dipole oscillation \cite{Pan2012} and roton modes
\cite{Roton1,Roton2}, as well as the dynamical instabilities~\cite{SOlattice}
in optical lattices, etc. Recently, the generation of another type of SO
coupling, the spin and orbital-angular-momentum coupling, was also proposed~%
\cite{Sun2015,Pu2015,Qu2015}.

Moving potential barriers have been used in the past for the study of the
superfluidity of ultra-cold atomic gases. For instance, by stirring a small
impenetrable barrier back and forth in a condensate, the evidence of the
critical velocity for a superfluid was observed \cite{Ketterle1999}. When a
wider and penetrable barrier was swept through a condensate at an
intermediate velocity, the condensate is filled with dark solitons \cite%
{Engels2007}.

In this paper, we study the moving barrier induced dynamics in a SO coupled
BEC. We find that a fast moving penetrable barrier may generate a dynamical
spin-density wave (SDW) in the wake of the barrier. Static SDW, which was
proposed in solid state physics by Overhauser~\cite{SDW1,SDW2}, has been
widely studied in many different solid state materials such as chromium~\cite%
{ReviewSDW,NatSDW}. Our generated SDW in a SO coupled BEC is induced by the
moving barrier and vanishes when the SO coupling is turned off. The spatial
periodic modulation of the spin density is not static, i.e., the local spin
polarization oscillates in time periodically, and could last for a very long
time.

The paper is organized as follows. Section~\ref{Sec:model} describes the
model of the SO coupled BEC. In Section~\ref{Sec:shock}, we study the
dynamics induced by a suddenly turned-on stationary barrier or a slowly
moving barrier. Section~\ref{Sec:SDW} includes the main results of the
paper. We generate a dynamical SDW with a fast moving barrier, study its
propagation, parameter dependence, and finally explain its mechanism using a
simple one-dimensional (1D) system. Section~\ref{Sec:conclude} is the
discussion.

\section{Theoretical model}

\label{Sec:model}

\begin{figure}[t]
\includegraphics[width=0.5\textwidth]{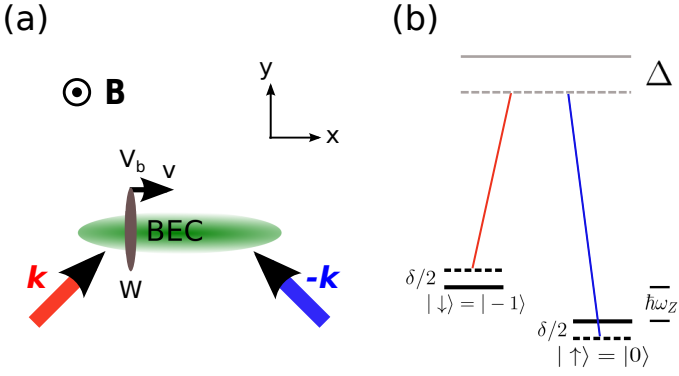} \centering
\caption{(Color online). (a) Experimental setup for the observation of
barrier induced dynamics in a SO coupled BEC. The realistic barrier, created
by another external laser beam, is characterized by the width $W$ (along $x$
direction), barrier peak potential $V_{b}$, and barrier velocity $v$. (b)
Energy level diagram of the SO coupled BEC. Two internal states are coupled
by the two-photon Raman transition through a pair of counter-propagating
Raman beams.}
\label{fig:setup}
\end{figure}

The SO coupled BEC is realized by shining two counter-propagating laser
beams on cold atoms \cite{Lin2011}. Two atomic internal states can be
regarded as pseudo-spins $|\uparrow \rangle =|F=1,m_{F}=0\rangle $ and $%
|\downarrow \rangle =|F=1,m_{F}=-1\rangle $ of $^{87}$Rb atoms. To stimulate
the two-photon Raman transitions, the two lasers are chosen to have a
frequency difference comparable with the Zeeman splitting $\hbar \omega _{Z}$
between two spin states. The experimental configuration and level diagram
are shown in Fig.~\ref{fig:setup}. In our simulations, we consider a
realistic elongated BEC with $N=10^{4}$ atoms in a harmonic trap with
trapping frequencies $\omega _{x,y,z}=2\pi \times \{20,120,500\}$ Hz. The
strong confinement along the $z$ and $y$ directions reduces the dimension to
quasi-1D. We use $E_{r}=\hbar ^{2}k_{L}^{2}/2m$ as the energy unit, where $%
k_{L}$ is the recoil momentum along the $x$ direction (i.e., the SO coupling
direction).

A barrier, which can be created by the dipole potential of another laser
beam~\cite{Ketterle1999,Engels2007}, is suddenly switched on in the BEC or
is swept from the left to the right side with a velocity of $v$ ranging from
$1$ to $80$ $\mu m/ms$. The barrier peak potential is $5$ to $25$ $E_{r}$,
which is much larger than the chemical potential of the system. The width of
the barrier is at the order of $\mu m$. The external potential barrier
sweeping through the BEC is modelled as a Gaussian potential of the form:
\begin{equation}
V(\mathbf{r},t)=V_{b}e^{-\frac{(x-x_{0}-vt)^{2}}{2w_{x}^{2}}-\frac{y^{2}}{%
2w_{y}^{2}}-\frac{z^{2}}{2w_{z}^{2}}},
\end{equation}%
where $w_{x}=W$ is the Gaussian barrier width along the SO coupling
direction. $w_{y/z}$ are much larger than the BEC widths in these two
directions. $x_{0}$ and $v$ are the initial position and velocity of the
barrier potential, $V_{b}$ is the peak potential of the barrier.

The dynamics of the SO coupled BEC are governed by the Gross-Pitaevskii (GP)
equation:
\begin{equation}
i\hbar \frac{\partial }{\partial {t}}\psi =[H_{SO}+V_{trap}+V(\mathbf{r}%
,t)+H_{I}]\psi ,  \label{G-P}
\end{equation}%
where the single-particle Hamiltonian with SO coupling is given by
\begin{equation}
{H_{SO}}=\frac{\hbar ^{2}}{2m}(\mathbf{k}+k_{L}\sigma _{z}\hat{e}_{x})^{2}+%
\frac{\delta }{2}\sigma _{z}+\frac{\Omega }{2}\sigma _{x},
\end{equation}%
where $\sigma _{i}$ $(i=x,y,z)$ are the Pauli matrices, $\delta $ is the
detuning of the Raman transition and $\Omega $ is the Raman coupling
strength. The trapping potential is of the form $V_{trap}=m\omega
_{x}^{2}x^{2}/2+m\omega _{y}^{2}y^{2}/2+m\omega _{z}^{2}z^{2}/2$. For the
sake of simplicity, we consider a 2D geometry in our calculations by
integrating out the $z$-dependent degree of freedom in the G-P equation \ref%
{G-P}, which is valid because the strong confinement along $z$ direction
restricts the BEC to the ground state of the harmonic trap along the $z$
direction, yielding
\begin{equation}
\psi _{3D}(\mathbf{r},t)=\psi _{2D}(x,y,t)\left( \frac{m\omega _{z}}{\pi
\hbar }\right) ^{1/4}e^{-\frac{m\omega _{z}}{2\hbar }z^{2}}.
\end{equation}

The interaction between atoms is determined by the mean-field Hamiltonian
\begin{equation}
H_{I}=\left(
\begin{array}{cc}
g_{\uparrow \uparrow }|\psi _{\uparrow }|^{2}+g_{\uparrow \downarrow }|\psi
_{\downarrow }|^{2} & 0 \\
0 & g_{\downarrow \uparrow }|\psi _{\uparrow }|^{2}+g_{\downarrow \downarrow
}|\psi _{\downarrow }|^{2}%
\end{array}%
\right) ,
\end{equation}%
where the reduced nonlinear coefficients for the 2D system are $g_{\uparrow
\uparrow }=\frac{2\sqrt{2\pi }\hbar ^{2}Nc_{0}}{ma_{z}}$, $g_{\uparrow
\downarrow }=g_{\downarrow \uparrow }=g_{\downarrow \downarrow }$=$\frac{2%
\sqrt{2\pi }\hbar ^{2}N(c_{0}+c_{2})}{ma_{z}}$. The harmonic oscillator
characteristic length is $a_{z}=\sqrt{\hbar /{m\omega _{z}}}$ and the $s$%
-wave scattering lengths are given by $c_{0}=100.86a_{0}$, $c_{2}=-0.46a_{0}$
($a_{0}$ is the Bohr radius).

\begin{figure}[t]
\includegraphics[width=0.5\textwidth]{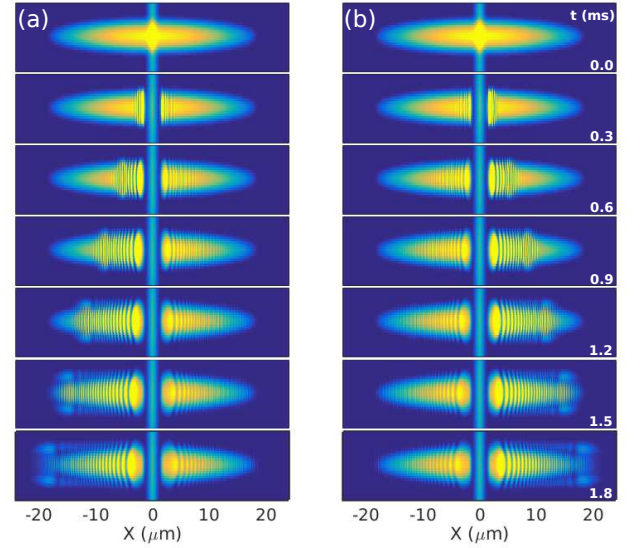} \centering
\caption{(Color online). Excitation and propagation of the spin-dependent
wave in a spin-orbit coupled BEC by suddenly switching on a stationary
barrier at the center of the trapped BEC at $t=0$ $ms$. Column (a) is for
spin up component, column (b) is for spin down component. $V_{b}=10E_{r}$, $%
W=0.5\protect\mu m$. Because of the presence of SO coupling, the propagation
of sound demonstrates an anisotropic behavior, i.e., spin up only propagates
to the left side while spin down only propagates to the right side.}
\label{fig:sound}
\end{figure}

In most cases the many-body interaction is weak, therefore many interesting
physics can be well understood from the single particle band structure,
which is either a double well type (for $\Omega <4E_{r}$) or a single well
type (for $\Omega >4E_{r}$) for $\delta =0$. For $\Omega <0.2E_{r}$, BEC
stays in both wells and the two dressed states interfere to form a stripe
pattern; for $0.2E_{r}<\Omega <4E_{r}$, BEC chooses either of the two wells
as the true ground state, which is usually called plane wave phase or
magnetized phase with a finite spin polarization $|\langle\sigma
_{z}\rangle|=\sqrt{1-\left( \Omega /4E_{r}\right) ^{2}}$; for $\Omega
>4E_{r} $, BEC condenses at $k_{x}=0$ and the spin polarization is zero $%
|\langle\sigma _{z}\rangle|=0$. In our calculations, we focus on the latter
case and take $\Omega =6E_{r}$, $\delta=0E_r$, where the generated SDW could
be identified easily.

One of the effects of the SO coupling is to change the sound of speed of the
condensate. For a regular BEC, the speed of sound is given by $v_{s}=\sqrt{%
U\rho /m}$, where $U=4\pi \hbar ^{2}Nc_{0}/m$ is the nonlinear coefficient
and $\rho $ is the condensate density. When the BEC is dressed by the Raman
lasers, the speed of sound is modified by changing the atomic mass to
effective mass $m_{eff}$ since the band structure of the system is modified,
i.e., $v_{s}=\sqrt{U\rho /m_{eff}}$. For experimentally relevant parameters,
the sound of speed is to the order of $1$ $\mu m/ms$. The sound of speed and
the collective excitation spectrum has been measured in recent SO coupling
experiments~\cite{Roton1,Roton2}. In order to generate SDW, the velocity of
the moving barrier should be much larger than the speed of sound.

\section{Effect of a suddenly turn-on stationary barrier and a slowly moving
barrier}

\label{Sec:shock}

\begin{figure}[t]
\includegraphics[width=0.5\textwidth]{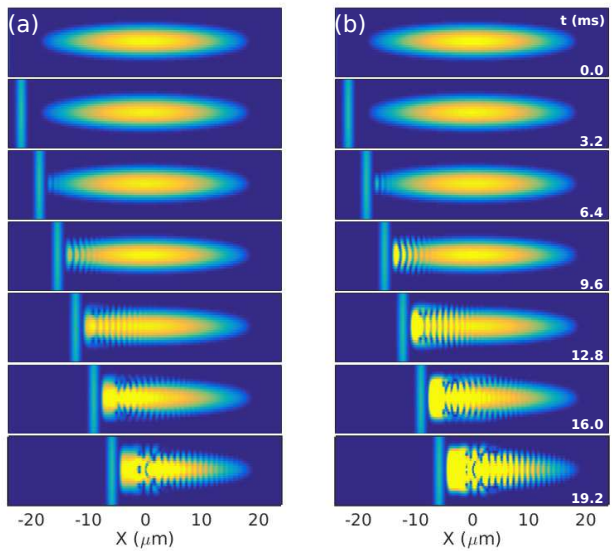} \centering
\caption{(Color online). A moving barrier with low velocity $v=1$ $\protect%
\mu m/ms$ pushes BEC to its right. $V_{b}=15E_{r}$, $W=0.5$ $\protect\mu m$.
Column (a) is for spin up component, column (b) is for spin down component.}
\label{fig:shock}
\end{figure}

Before discussing the generation of SDW in the SO coupled BEC through a fast
moving barrier, we consider two different limits: one is that a stationary
barrier is suddenly switched on in the middle of the condensate, the other
one is that a slowly moving barrier is swept through the BEC from the left
side to the right side. Previously, suddenly switching on a barrier
potential in the middle of the superfluid is usually used to measure the
speed of sound for BECs or Fermi gases~\cite{sound1,sound2}. In our
calculations, the barrier potential is strong and therefore induces strong
perturbations to the condensate which might be observed in experiments.
Without SO coupling, the barrier excites two wave fronts propagating along
both directions with the same speeds for the two spins. For SO coupled BEC,
as shown in Fig.~\ref{fig:sound}, the propagation of the two wave fronts
propagate differently with respect to the direction of the spin. We see that
only one clear wave front propagates to the left (right) side for spin up
(down) with a speed around $\sim 10$ $\mu m/ms$ for the current geometry and
atom number. This anisotropic and spin-dependent propagation of the density
perturbation is a direct consequence of the SO coupling. Note that except
the propagation of the wave fronts, a series of density modulations are
excited in the mean time, which are due to the suddenly switched on barrier
induced perturbations and occur for a regular BEC as well.

When the barrier is slowly swept from the left side to the right side of the
BEC, it is impenetrable for the condensate because the barrier potential $%
V_{b}=15E_{r}$ is much larger than the chemical potential $\sim 1$ $E_{r}$
of the condensate. Therefore BEC is pushed in front of the barrier with the
excitations of similar density modulations for the two spins (see Fig. \ref%
{fig:shock}).

\section{SDW from a fast moving barrier}

\label{Sec:SDW}

In this section, we focus on a fast moving barrier with a velocity larger
than the speed of sound. Because of the increasing relative velocities
between the barrier and the condensate, the barrier is now a penetrable
potential for the atoms. For a certain parameter regime, the barrier induces
a dynamical modulation of the densities of the two spins in the wake of the
barrier, while it does not lead to any observable perturbations in its front.

\subsection{Generation and propagation of SDW}

\begin{figure}[t]
\includegraphics[width=0.5\textwidth]{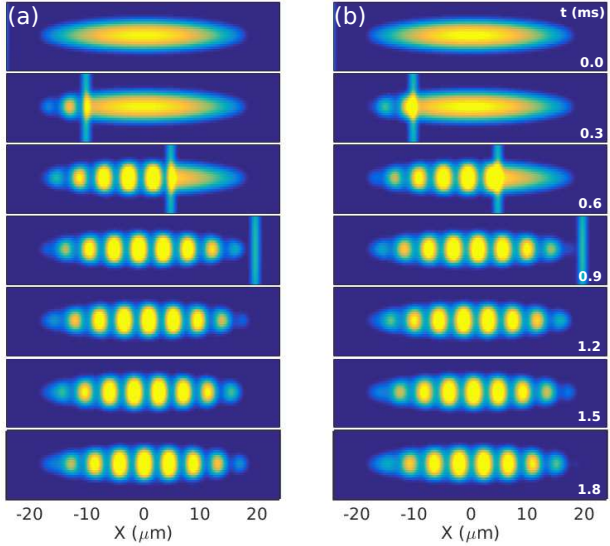} \centering
\caption{(Color online). A fast moving barrier with a velocity of $v=50%
\protect\mu m/ms$ penetrates the BEC and generates a dynamical SDW. $%
V_b=15E_r$, $W=0.5\protect\mu m$. Column (a) is for spin up component,
column (b) is for spin down component.}
\label{fig:SDW-gen}
\end{figure}

\begin{figure}[t!]
\includegraphics[width=0.5\textwidth]{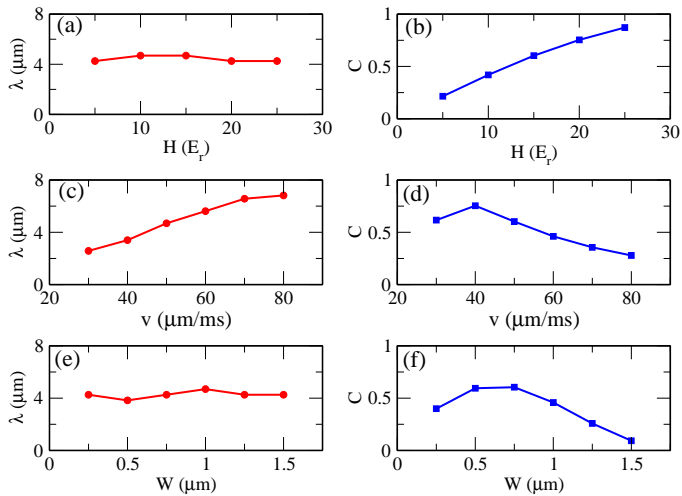} \centering
\caption{(Color online). The wavelength $\protect\lambda$ and the contrast $%
C $ of the SDW generated in a fast moving barrier as a function of (a,b) the
barrier peak potential $V_b$ for $v=50\protect\mu m/ms$, $W=0.5\protect\mu m$%
, (c,d) the barrier velocity $v$ for $V_b=15E_r$, $w=0.5\protect\mu m$,
(e,f) the width $W$ for $v=50\protect\mu m/ms$, $V_b=15E_r$. }
\label{fig:SDW-period}
\end{figure}

\begin{figure}[t]
\includegraphics[width=0.5\textwidth]{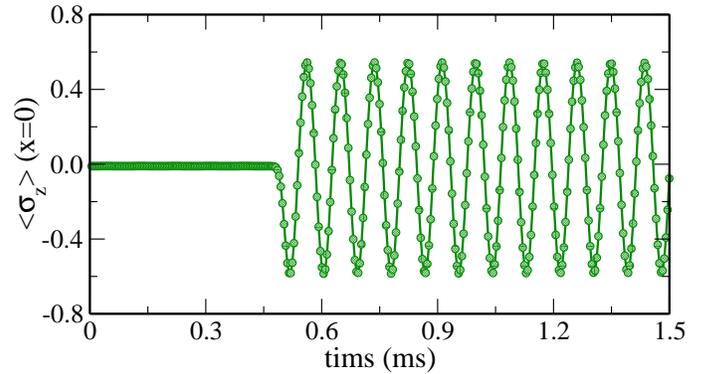} \centering
\caption{(Color online). Local spin polarization $\langle \protect\sigma %
_{z}\rangle (x=0)$ as a function of time for Fig.~\protect\ref{fig:SDW-gen}.}
\label{fig:SDW-time}
\end{figure}

Figure~\ref{fig:SDW-gen} shows the density distributions of two spin
components when and after a moving barrier with a fast velocity is swept
through the condensate. The velocity of the barrier $v=50$ $\mu m/ms$ is
much larger than the speed of sound. We see that the density oscillations of
the two spin components are out of phase, thus the barrier generates a SDW.
The SDW could last a very long time and does not relax in the trap if all
the parameters remain unchanged.

To characterize the SDW, we calculate the wavelength $\lambda $ (the
distance between two peaks for one spin component) and the contrast $%
C_{\sigma }=|n_{max}^{\sigma }-n_{min}^{\sigma }|/|n_{max}^{\sigma
}+n_{min}^{\sigma }|$ near the center of the BEC and plot them as a function
of the barrier's peak potential $V_{b}$, barrier's velocity $v$ and
barrier's width $W$. As shown in Fig.~\ref{fig:SDW-period}, the wavelength
of the SDW is roughly a constant as a function of barrier height and width.
However, the wavelength is almost proportional to the velocity of the
barrier.

This is easy to understand because the SDW is not static. Each local spin
polarization is fast oscillating as a function of time, as shown in Fig.~\ref%
{fig:SDW-time} where we plot the local spin polarization $\langle \sigma
_{z}\rangle (x=0)$ in the middle of the BEC as a function of time. Before
the barrier moves to $x=0$, the density is a constant. Right after the
barrier is swept through $x=0$, the densities of the two spins start \textit{%
out-of-phase} oscillations, therefore there is a spin polarization
oscillation. The oscillation period is roughly a constant $T$ for a certain
barrier potential $V_{b}$. Considering that the barrier is swept through the
BEC, the left and right side of the BEC is perturbed consecutively. When the
barrier moves at a constant speed, the wavelength of the SDW should be
proportional to the moving velocity if $T$ does not depend on the velocity
significantly, i.e., $\lambda =v\times T$. Note that this relation fails to
apply for much larger velocities where the spin oscillation at different
local points may be generated at very short time and the above physical
picture does not apply. In our numerical calculations, we verify that this
oscillation curve remains the same when the interaction strength is varied
in a large region, showing that the dynamical SDW is a phenomena governed by
the single-particle physics. However, the phenomena changes when the
interaction strength is strong enough such that the speed of sound of the
condensate is comparable to the moving barrier velocity. As we have studied
previously in Fig. \ref{fig:shock}, the barrier becomes impenetrable in this
limit and pushes BEC to one side.

\begin{figure}[t]
\includegraphics[width=0.5\textwidth]{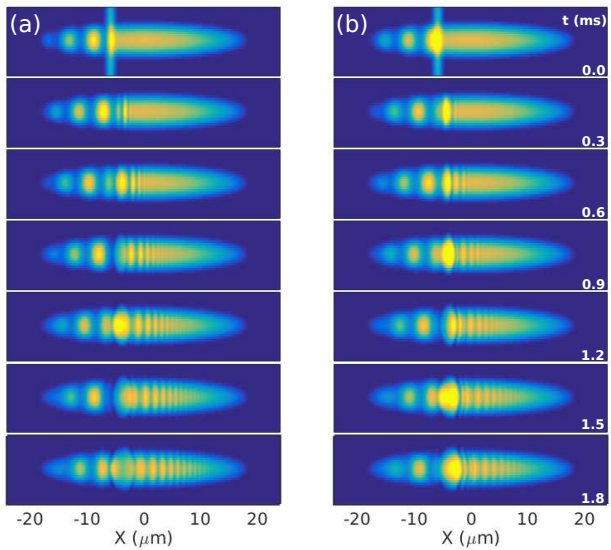} \centering
\caption{(Color online). Propagation of SDW in a SO coupled BEC. When the
barrier moves to the center of the BEC, we suddenly switch off the barrier
and observe the slow propagation of the SDW to the unperturbed region. All
the parameters are the same as Fig.~\protect\ref{fig:SDW-gen}.}
\label{fig:SDW-prop}
\end{figure}

In Fig.~\ref{fig:SDW-prop}, we demonstrate the slow propagation of the SDW
when the barrier is suddenly turned off after it moves to the center of the
BEC. Since the dynamical SDW is actually a local spin polarization
oscillation instead of travelling waves, the removal of the moving barrier
at the center of the BEC should stop generating SDW on its right side.
However, because of the superfluidity properties of the BEC and nonlinear
interactions, the unperturbed neighboring atoms will be eventually perturbed
and we see that a new SDW with much smaller amplitude and velocity
propagates to the right side of the BEC. The propagation of the new SDW is
in the order of the speed of sound as expected.

\subsection{Mechanism of SDW generation}

In this section, we provide a qualitative theoretical understanding of the
SDW generation by considering a rectangular potential sweeping through a 1D
SO coupled BEC system. The rectangular potential can be written as
\begin{equation}
V(x,t)=V_{b}[\mathbb{H}(vt-x_{0}-W)-\mathbb{H}(vt-x_{0})],
\end{equation}%
where $\mathbb{H}(x)$ is Heaviside function:
\begin{equation}
\mathbb{H}(x)=\left\{
\begin{array}{ll}
0 & \text{$x<0$} \\
1 & \text{$x\geq 0$}%
\end{array}%
\right. .
\end{equation}%
Similar as the Gaussian barrier, here $V_{b}$ is the barrier potential and $W
$ is the barrier width.

\begin{figure}[t]
\includegraphics[width=0.5\textwidth]{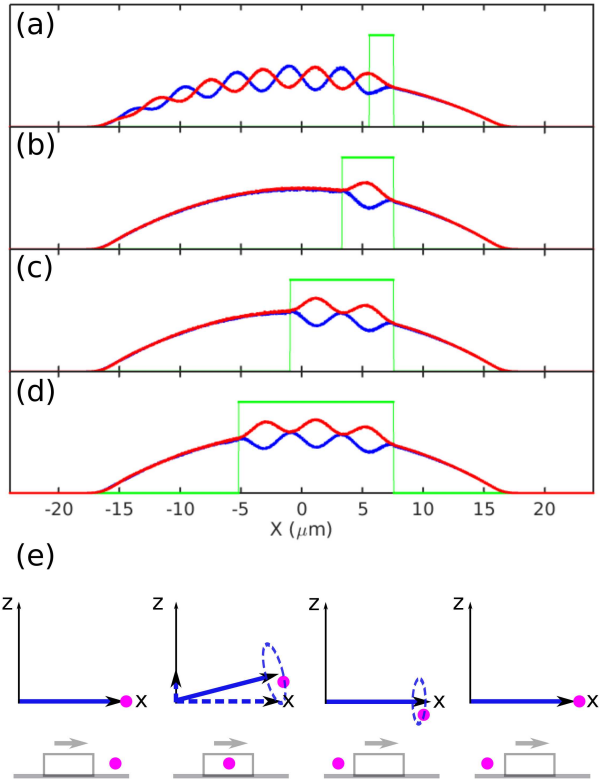} \centering
\caption{(Color online). SDW generation from a rectangular barrier in 1D SO
coupled BEC system for different barrier widths (a) $W=2\protect\mu m$ (b) $%
W=4.25\protect\mu m$, (c) $W=8.5\protect\mu m$, (d) $W=12.75\protect\mu m$.
Other parameter are $V_{b}=4E_{r}$, $v=50\protect\mu m/ms$. (e) The dynamics
of the system is equivalent to the precession of a spin under the magnetic
field. Without barrier, the system is in the eigenstate of a horizontal
magnetic field $B_{x}$. A moving barrier is equivalent to suddenly quenching
on a $z$ component magnetic field $B_{z}$. The spin then does precession
about the axis determined by the total magnetic field. When the barrier
passes the point, the spin could be in any possible orientations (with SDW
appears) or returns to the eigenstate of $B_{x}$ (no SDW appears). Other
parameters are: $\Omega =6E_{r}$, $\protect\delta =0E_{r}$, $\protect\omega %
_{x}=20Hz$.}
\label{fig:SDW-1D}
\end{figure}

Figure~\ref{fig:SDW-1D} shows density distributions of the two spins at the
moment when the rectangular barrier moves to a position around $X=8\mu m$
for different barrier widths. We see that the width of the barrier changes
the generated SDW significantly. When $L$ is integer times larger than the
SDW wavelength $L=n\lambda $, there are only $n$ well developed complete SDW
oscillations inside the barrier and the BEC that behind the barrier seems to
be unperturbed at all (Fig.\ref{fig:SDW-1D}b,c,d). When $L$ and $\lambda $
are incommensurate, the SDW is generated in the wake of the barrier as shown
in Fig.~\ref{fig:SDW-1D}a. The wavelength of the SDW is also proportional to
the velocity as we have demonstrated in Fig.~\ref{fig:SDW-period}(c) for the
Gaussian-shaped barrier, while rarely depends on the barrier width and
height. All these features of the dynamics can be well understood in the
following way.

The condensate is initially prepared at the ground state with $\Omega =6E_{r}
$ while the band structure has only one single minimum at $k_{x}=0$, i.e.,
the quasi-momentum of the SO coupled BEC is zero. The initial wave function
at some point $x$ is
\begin{equation}
\psi (x,t=0)=\frac{1}{\sqrt{2}}\left(
\begin{array}{c}
1 \\
-1%
\end{array}%
\right) .
\end{equation}

The effect of a moving barrier could be explained using a simple single
particle picture. In the laboratory frame, the real momentum of spin up and
spin down components is $k_{\uparrow }=k_{x}+k_{L}=k_{L}$ and $k_{\downarrow
}=k_{x}-k_{L}=-k_{L}$ respectively because $k_{x}=0$. An external barrier
that moves along the SO coupling direction has different relative velocities
for the two spins and therefore induces spin-dependent dynamics. Take the
traveling external potential as the frame of reference, the momentum are
then $k_{\uparrow B}=-mv/\hbar +k_{L}$ for spin up and $k_{\downarrow
B}=-mv/\hbar -k_{L}$ for spin down. In the presence of the fast moving
barrier, the velocities of the two spin components will be changed. From the
conservation of the energy, we have the new velocities for the two spins: $%
k_{\uparrow B}^{\prime }=-\sqrt{k_{\uparrow B}^{2}-k_{B}^{2}}$ and $%
k_{\downarrow B}^{\prime }=-\sqrt{k_{\downarrow B}^{2}-k_{B}^{2}}$ where $%
k_{B}=\sqrt{2mV_{b}/\hbar ^{2}}=2k_{L}$ since $V_{b}=4E_{r}$. Now converting
to the laboratory frame, the velocity is $k_{\uparrow }^{\prime
}=k_{\uparrow B}^{\prime }+mv/\hbar $ for spin up, and $k_{\downarrow
}^{\prime }=k_{\downarrow B}^{\prime }+mv/\hbar $ for spin down. We find
that for the large barrier velocity, $k_{\uparrow }^{\prime }-k_{\downarrow
}^{\prime }\approx 2k_{L}$, which means the quasi-momentum for this states
is now $k_{x}^{\prime }=(k_{\uparrow }^{\prime }+k_{\downarrow }^{\prime })/2
$ that agrees with the appearance of new momentum states in our GP
simulations. Because of different group velocities of two spins in the
presence of SO coupling, the moving barrier drives BEC to a new nonzero
quasi-momentum states which is equivalent to an effective detuning $\Delta
_{e}$.

Now the SDW related phenomena could be modelled as a quench dynamics where
the effective detuning $\Delta _{e}$ is suddenly added to induce a coupling
between the two new bands. Consider the effective Hamiltonian for a
two-level system (ignoring other irrelevant constants):
\begin{equation}
H_{eff}=\left(
\begin{array}{cc}
\Delta _{e}/2 & \Omega /2 \\
\Omega /2 & -\Delta _{e}/2%
\end{array}%
\right) .  \label{eq:eps}
\end{equation}%
We denote $\Omega _{e}=\sqrt{\Omega ^{2}+\Delta _{e}^{2}}$, then the
evolution of the local wave function for the point $x_{0}$ within the
potential is given by:
\begin{equation}
\psi (x_{0},t)=\frac{1}{2\sqrt{2}(1-\frac{\Delta _{e}}{\Omega })}\left(
\begin{array}{c}
2-\frac{3\Delta _{e}}{\Omega }+\frac{\Delta _{e}}{\Omega }e^{i\Omega _{e}t}
\\
-2+\frac{\Delta _{e}}{\Omega }+\frac{\Delta _{e}}{\Omega }e^{i\Omega _{e}t}%
\end{array}%
\right) .
\end{equation}

Ignoring the normalized factor and the small terms of the order $\Delta
_{e}^{2}$, a straightforward calculation gives the spin polarization at this
local point,
\begin{equation}
\langle \sigma _{z}(x_{0},t)\rangle =C_{1}\frac{\Delta _{e}}{\Omega }[\cos
(\Omega _{e}t)-1],
\end{equation}%
where $C_{1}>0$.

It is quite clear that when the barrier moves to $x_{0}$ with a fast
velocity, it perturbs the local condensate $n_{\sigma }(x_{0},t)$ by
coupling the two new bands and thus induces a \textit{local} spin
polarization oscillation $\langle \sigma _{z}(x_{0},t)\rangle $. Due to the
fact that the perturbation is applied from left to right, there is a
relative phase between neighboring atoms. Therefore the spin polarization
for a point on the left of $x_{0}$ is
\begin{equation}
\langle \sigma _{z}(x,t)\rangle =C_{1}\frac{\Delta _{e}}{\Omega }[\cos
(\Omega _{e}t-\frac{x-x_{0}}{v})-1].
\end{equation}%
We see from the above equation that the local spin polarization within the
rectangular potential is always negative and the period of the oscillation
is $T_{in}=2\pi \hbar /\Omega _{e}$. Therefore the wavelength is $\lambda
_{in}=2\pi \hbar v/\Omega _{e}$.

Note that the point which has a separation of $n\lambda _{in}$ with the
right edge of the rectangular barrier has a vanishing spin polarization. If
the left edge of the barrier coincides with these spin polarization
vanishing points, i.e., $W=n\lambda _{in}$, then a striking effect occurs as
we have seen from Fig.~\ref{fig:SDW-1D}b-d: the condensate in the wake of
the barrier seems to be unperturbed at all. That is because the potential
has been removed for these points, where the dynamics is now governed by the
original Hamiltonian with $\Delta _{e}=0$. At the same time the wave
function returns to its eigenstate with $\langle \sigma _{z}\rangle =0$.

If $W\neq n\lambda _{in}$, then even though the governed Hamiltonian returns
to the original one, the wave function is not in its eigenstate and the
coupling between the old two bands continues with the new Rabi frequency.
Therefore in the wake of the barrier, we have the period $T_{out}=2\pi \hbar
/\Omega >T_{in}$ and the wavelength $\lambda _{out}=vT_{out}$ which is a
little larger than $\lambda _{in}$ because $\Omega _{e}$ is slightly larger
than $\Omega $. Similar calculation shows that the spin polarization could
be positive or negative in one period, in agreement with the GP simulation
(Fig.~\ref{fig:SDW-1D}a). Furthermore, when $W=(2n+1)\lambda _{in}/2$, there
is the largest spin polarization oscillation amplitude behind the barrier.

The dynamics of the system are equivalent to the precession of a spin under
the magnetic field (Fig.~\ref{fig:SDW-1D}e). Without the barrier, the system
is in the eigenstate of a horizontal magnetic field $B_{x}$ and thus does
not precess. A moving barrier is equivalent to suddenly quenching on a $z$
component magnetic field $B_{z}$. The spin then precesses about the axis
determined by the total magnetic field. When the barrier passes through the
point, the spin could be in any possible orientations (with SDW appears) or
returns to the eigenstate of $B_{x}$ (no SDW appears).

\section{Discussion}

\label{Sec:conclude}

In our calculation, we have chosen $\Omega =6E_{r}$. Similar SDWs can also
be generated for $\Omega <4E_{r}$, where the initial state is polarized and
the population oscillation amplitudes of the two spins are now different. If
the barrier is even faster than the speed used in this paper, then all atoms
are perturbed almost at the same time, therefore the local spin polarization
oscillation may be in phase. Since the delay is always present no matter how
small it is, the pattern may look complicated. For a two-component BEC
without SO coupling, a fast moving barrier does not induce any observable
effects because the dynamics is governed by two uncoupled bands.

The two-photon recoil momentum and recoil energy correspond to a velocity of
$v_{r}=\hbar k_{L}/m=4.14\mu m/ms$ and a kinetic energy of $1E_{r}$.
According to our simulations, to generate the SDW with a fast moving barrier
(velocity $v$, peak potential $V_{b}$), we need to focus on the following
parameter regime:
\begin{equation}
KE_{b}>>V_{b}>>E_{r},
\end{equation}%
where $KE_{b}=\frac{1}{2}mv^{2}$ is the corresponding kinetic energy of a
particle with a relative speed of $|\pm v_{r}-v|\approx v$ (because $v>>v_{r}
$) respect to the barrier.

In summary, we present a scheme to observe the generation and propagation of
a SDW in a SO coupled BEC through a moving supersonic potential. The period
of the SDW almost does not change with respect to the peak potential and
width of the barrier. However, it is very sensitive to the velocity of the
barrier. The essence of the SDW is due to the different group velocities of
the two spin components in the presence of SO coupling. The SDW could last a
long time in the trap without relaxation and therefore provides a good
system to study other complicated dynamics. For instance, by lowering the
Raman coupling and changing the band structure, we may observe the opposite
motion of the density modulations of the two spins and their relaxation in
the presence of SO coupling. Furthermore, in other parameter regime (smaller
barrier potential or velocity) or with a much narrower stirring barrier, it
is possible to generate solitons or vortices. \newline

\begin{acknowledgments}
We thank Peter Engels for useful discussions. Y. Li is supported by NSFC
under Grant No. 11104075, and the China Scholarship Council.  C. Qu and C. Zhang are supported by ARO (W911NF-12-1-0334) and AFOSR
(FA9550-13-1-0045). Y. Zhang is
supported by NSFC(61275122), the National Fundamental Research
Program(2011CB921200, 2011CBA00200), K. C. Wong Education Foundation and
CAS.
\end{acknowledgments}


\end{document}